\documentclass[twocolumn,a4paper]{article}

\usepackage{bm} % bold math
\usepackage{amssymb,amsfonts}
\usepackage{abstract}
\usepackage{color}
\usepackage[pdftex,urlcolor=blue]{hyperref}

\begin{document}

\title{\bf{\huge{Reverse Engineering Approach\\to Quantum Electrodynamics}}}

\author{\bf{Walter Smilga} \\ Isardamm 135 d, 82538 Geretsried, Germany \\ Email:\,\,wsmilga@compuserve.com }

%\date{Received 18.02.2013}

\twocolumn[
  \begin{@twocolumnfalse}
    \maketitle
\begin{abstract}

The $S$ matrix of $e$--$e$ scattering has the structure of a projection operator that 
projects incoming separable product states onto entangled two-electron states. 
In this projection operator the empirical value of the fine-structure constant 
$\alpha$ acts as a normalization factor.
When the structure of the two-particle state space is known, a theoretical 
value of the normalization factor can be calculated.
For an irreducible two-particle representation of the Poincar\'e group, the 
calculated normalization factor matches Wyler's semi-empirical formula for the 
fine-structure constant $\alpha$.
The empirical value of $\alpha$, therefore, provides experimental evidence that 
the state space of two interacting electrons belongs to an irreducible 
two-particle representation of the Poincar\'e group.
\end{abstract}


Parts of this article were presented at the 7th International Conference On Quantum Theory and Symmetries (QTS7) in Prague 2011 \cite{smi12}.

\vspace{1cm}

\end{@twocolumnfalse}
]

\section{Introduction}

The development of quantum electrodynamics (QED) belongs to the 
greatest successes of theoretical physics. 
Provided that a sufficient number of terms of the perturbation series are 
included, the results of QED agree with the experimental data to any required 
degree of precision.
This is a strong support for the correctness of the perturbation algorithm of 
QED.
Nevertheless, we are far from completely understanding this algorithm.
Although the success of QED has widely been considered as a confirmation 
of the concept of interacting quantum fields, i.e., of the electron field's 
interacting with the photon field, theoretical considerations (e.g., Haag's 
Theorem \cite{rh55}) call into doubt that QED is really a quantum field theory 
of interacting fields. 
Aside from this open question of the compatibility of QED with the concepts of 
quantum field theory, notorious divergences plague the users of the algorithm.
These divergences can be removed by renormalization, but their mere
existence makes it difficult to really understand the perturbation algorithm. 
This does not prevent the majority of practitioners of QED from successfully 
using the perturbation algorithm, following the famous slogan: 
``Shut up and calculate'' \cite{ndm}.

A similar situation is often encountered in software engineering, when a 
software program is available only as a (machine readable) object program, 
but not as (human readable) source code.
Here, such situations are successfully handled by means of ``reverse 
engineering'' \cite{ee}. 
From Wikipedia \cite{wikire}:
``Reverse engineering is the process of discovering the technological 
principles of a device, object, or system through analysis of its structure, 
function, and operation.'' 

The term ``reverse engineering'' originally described the (sometimes illegal) 
use of mechanical engineering to analyze competitor's products, when the 
original blue prints, for understandable reasons, were not available.
Nowadays, reverse engineering is well-known in software engineering as
a powerful, though sometimes cumbersome, method for reconstructing the original 
source code of a program by decompiling or disassembling the binary 
machine code when the source code is not available -- whether it has been 
lost or whether it has not been made available by the original manufacturer. 

When we buy a software product, we usually have to sign a licensing agreement 
similar to: ``The use of the software is subject to the following 
restrictions: You are prohibited from decompiling, reverse engineering, or 
disassembling the software, or otherwise attempting to derive their source 
code.''
In QED we are in the advantageous position that its perturbation algorithm is 
``public domain,'' although we are not sure whether or not we are in the 
possession of the correct and complete ``source code''. 
In any case, there is no licensing agreement that can prevent us from 
reconstructing the ``source code'' by reverse engineering.
In view of six decades of ``Shut up and calculate,'' at least an attempt 
is long overdue.

In line with the approach used in software engineering, we will isolate the basic 
building blocks of the perturbation algorithm, and find each one's mathematical 
functionality.
Then we will put these building blocks together, to find their combined 
functionality.
If carefully done, this will result in a consistent description of the 
perturbation algorithm, which can be regarded as the ``source code'' behind the
algorithm.
This description may then serve as a basis for a physical interpretation. 
It should not come as a surprise, however, if this interpretation turns 
out to not reproduce the physical concepts that historically led to the design of 
the perturbation algorithm.

Reverse engineering is usually followed by re-engineering the object under
study, with the goal of improving or extending its functionality.
The present paper is limited to the reverse engineering phase, and we will 
take strict care not to change the perturbation algorithm.

\section{A short review of quantum\\electrodynamics}

The following is a short overview of QED, as formulated by Feynman in his 
seminal papers of 1949/50 \cite{rf1,rf2,rf3}.
 
QED uses a perturbation approach to the $S$ matrix, which, for an electromagnetic 
scattering process, delivers the transition probabilities between the incoming and 
outgoing two-particle states.
The incoming and outgoing states are described by states in Fock space.
These states are constructed through repeated application of ``creation'' 
operators to a ``vacuum'' state.
A particle in a Fock state can be annihilated through a corres\-ponding 
``annihilation'' operator.
Creation and annihilation operators satisfy certain commutation 
or anticommutation rules, which ensure that the generated 
multi-particle states have the correct symmetry of either Fermi--Dirac
statistics (electrons) or Bose--Einstein statistics (photons).
Multi-particle states are first generated as pure product states.
They are used to describe the ``incoming'' and ``outgoing'' states. 
Because these states are separable, there are no correlations between the 
individual particle states other than by the mentioned statistics, so that 
the incoming and outgoing states describe ``free'' particles.
Linear combinations of separable product states, which in general will not
be separable but entangled, then make up a full product state 
space, corresponding to a product representation of the 
Poincar\'e group. 

The idea behind the concept of the $S$ matrix is that without knowing 
exactly what happens in the ``interaction region,'' we should formulate a 
quantum mechanical scattering theory on the basis of the incoming and outgoing 
states, because only these states are directly accessible to the 
experimenter \cite{wh}.
But since the incoming and outgoing states describe non-interacting 
particles, a heuristic ``interaction term'' is needed, to describe, at 
least in a phenomenological form, the process inside the interaction region.
Since it seems reasonable that the interaction process is uniquely determined 
by incoming and outgoing states, it has been tried to construct interaction 
terms from creation and annihilation operators of the incoming and outgoing 
states.
Relativistic (Poincar\'e) invariance greatly restricts the structure of such 
terms.
It turns out that with the additional requirement of gauge invariance
(of second kind), the interaction term
\begin{equation}
e\; \bar{\psi}(x)\, \gamma^\mu \, \psi(x)\, A_\mu(x)  \label{1-1}
\end{equation}
is uniquely determined, up to a constant factor $e$.
The factor $e$, the {\it electromagnetic coupling constant}, has been 
determined experimentally.
Its square is the {\it electromagnetic fine-structure constant} $\alpha$
(with the convention $\hbar = c = 1$).
The {\it field operators} $\bar{\psi}(x)$, $\psi(x)$ and $A_\mu(x)$ are 
operator-valued distributions.

$\bar{\psi}(x)$ and $\psi(x)$ are field operators of the electron--positron 
field (cf.\,\,e.g.\,\,Scharf \cite{sch89})
\begin{eqnarray}
&&\hspace{-0.5cm}\psi(x) = \label{1-2} \\
&&\hspace{-0.5cm}(2\pi)^{-3/2}\!\!\!\int\!\!d{\mathbf{p}}\!
\left( 
          b_s({\mathbf{p}}) u_s({\mathbf{p}}) e^{-ipx}  
\!\!+\!  d_s({\mathbf{p}})^\dagger v_s({\mathbf{p}}) e^{ipx} \! 
\right), \nonumber
\end{eqnarray} 
$\bar{\psi}(x) =\psi(x)^\dagger \gamma^0$ is the Dirac adjoint operator,
$\gamma^\mu$ are the Dirac matrices, and $\dagger$ means Hermitian adjoint. 
$u_s({\mathbf{p}})$ and $v_s({\mathbf{p}})$ are solutions of the Dirac 
equation of, respectively, positive and negative energy.

$A_\mu(x)$ is the field operator of the electromagnetic field 
\begin{eqnarray}
&&\hspace{-0.5cm}A_\mu(x) =   \label{1-3} \\
&&\hspace{-0.5cm}(2\pi)^{-3/2} \!\!\int\!\! 
\frac{d{\mathbf{k}}}{\sqrt{2 k_0}} 
\left( a_\mu({\mathbf{k}}) e^{-ikx} 
+ a_\mu({\mathbf{k}})^\dagger e^{ikx} \right)   \nonumber    
\end{eqnarray} 
(ignoring the fact that $A_0(x)$ is usually defined in a slightly 
different way to ensure manifest Lorentz covariance).

The creation operator $b_s({\mathbf{p}})^\dagger$ creates from the ``vacuum
state'' $\left|0\right>$ an electron state with momentum ${\mathbf{p}}$ and 
spin $s$, 
$\left|{\mathbf{p},s}\right> = b_s({\mathbf{p}})^\dagger \left|0\right>$.
The Hermitian adjoint operator $b_s({\mathbf{p}})$ is the corresponding 
annihilation operator; 
for the vacuum state $b_s({\mathbf{p}})\left|0\right> = 0$ holds.
$d_s({\mathbf{p}})^\dagger$, $d_s({\mathbf{p}})$ are the respective 
operators for positrons.
$a_\mu({\mathbf{k}})^\dagger$, $a_\mu({\mathbf{k}})$ 
create and annihilate a photon with momentum ${\mathbf{k}}$.
We have the anticommutation rules 
\begin{eqnarray}
\left\{ b_s({\mathbf{p}}),\, b_{s'}({\mathbf{p'}})^\dagger \right\} 
\!\!\!&\equiv&\!\!\!b_s({\mathbf{p}})\, b_{s'}({\mathbf{p'}})^\dagger 
+ b_{s'}({\mathbf{p'}})^\dagger b_s({\mathbf{p}}) \nonumber  \\
\!\!\!&=&\!\!\!\delta_{s s'}\,\delta^3({\mathbf{p - p'}}) \; , \label{1-4a} \\
\left\{ b_s({\mathbf{p}}),\, b_{s'}({\mathbf{p'}}) \right\}\!\!\!&=&\!\!\! 
\left\{ b_{s}({\mathbf{p}})^\dagger, \, b_{s'}({\mathbf{p'}})^\dagger\right\}
= 0   \label{1-4b}
\end{eqnarray}
-- analogous rules apply to $d_{s}({\mathbf{p}})^\dagger \mbox{ and }
d_{s}({\mathbf{p}})$ -- and the commutation rules
\begin{eqnarray}
\left[ a_\mu({\mathbf{k}}), \, a_\nu({\mathbf{k'}})^\dagger \right]
\!\!\!&\equiv&\!\!\!a_\mu({\mathbf{k}})\, a_\nu({\mathbf{k'}})^\dagger 
- a_\nu({\mathbf{k'}})^\dagger a_\mu({\mathbf{k}})	\nonumber     \\
\!\!\!&=&\!\!\!\delta_{\mu\nu}\,\delta^3({\mathbf{k-k'}}) \;,	\label{1-4c}	\\
\left[ a_\mu({\mathbf{k}}), \, a_\nu({\mathbf{k'}}) \right]\!\!\!&=&\!\!\! 
\left[ a_\mu({\mathbf{k}})^\dagger, \, a_\nu({\mathbf{k'}})^\dagger \right]	= 0.
\label{1-4d}									
\end{eqnarray}

The lack of precise information about the ``physical'' processes inside the 
interaction region, and the association of the terms ``creation'' and 
``annihilation'' with real dynamic processes, has led to our present picture 
of QED:
a highly dynamic, not to say chaotic, interplay of particles, 
continuously created from the vacuum, annihilated just a short time later, only 
controlled by some conserved quantum numbers, such as charge and lepton number.

\section{The S matrix of (elastic)\\electron--electron scattering}

The perturbation approach to QED uses the interaction term (\ref{1-1}) as a
``perturbation'' to the ``free'' theory and expands the $S$ matrix into a 
series of increasing orders in $e^2$.
The first order contribution is obtained from the two-point distribution 
built from an iteration of the interaction term,
\begin{eqnarray}
D_2(x_1,x_2) =
e^2\!\! &:&\!\!\!\bar{\psi}(x_1)\, \gamma^\mu \, \psi(x_1): A_\mu(x_1) \nonumber \\
     \!\!&:& \!\!\!\bar{\psi}(x_2)\, \gamma^\nu \, \psi(x_2): A_\nu(x_2), \label{1-12}
\end{eqnarray}
where the colons ``$:\dots:$'' mean ``normal ordering'' (cf.\,\,e.g.\,\,Scharf 
\cite{sch89}).
Higher orders are constructed by iterating this first order contribution.

After inserting the explicit form of the field operators (\ref{1-2}) and 
(\ref{1-3}) into the two-point distribution (\ref{1-12}), the corresponding 
first-order $S$ matrix, 
\begin{equation}
S_{12} = \int d^4x_1 d^4x_2 \,D_2(x_1,x_2) 
\label{1-5}
\end{equation}
can be evaluated.
By combining the phase factors of the field operators(\ref{1-2}) and 
(\ref{1-3}) with the integrations in equation (\ref{1-5}), we can construct 
$\delta$ functions of the form
\begin{equation}
\frac{1}{(2\pi)^4} \int \! d^4x \, e^{ix(p - k)}	= \delta^4(p - k)  ,	
\label{1-13}
\end{equation}
which can be used to rearrange the momenta.
As an intermediate result, we obtain several terms of the structure 
(all c-numbers are replaced by ``$\dots$'')
\begin{eqnarray}
\int\!\! 
d{\mathbf{p_1}}d{\mathbf{p_2}}d{\mathbf{k_1}}d{\mathbf{k_2}} &\dots&
b^\dagger({\mathbf{p_1+k_1}})\, \gamma^\mu \, b({\mathbf{p_1}}) \nonumber \\
&\times&b^\dagger({\mathbf{p_2-k_2}})\, \gamma_\nu \, b({\mathbf{p_2}}) \nonumber \\
&\times&a_\mu({\mathbf{k_1}})\,a_\nu^\dagger({\mathbf{k_2}}) .
\label{1-6}
\end{eqnarray}

Contraction (permutation) of the photon operators results in  
$\delta_{\mu\nu}\,\delta^3({\mathbf{k_1-k_2}})$.
By integrating over ${\mathbf{k_1}}$, we obtain
\begin{eqnarray}
\int 
d{\mathbf{p_1}}d{\mathbf{p_2}}\,d{\mathbf{k}} &\dots&
b^\dagger({\mathbf{p_1+k}})\, \gamma^\mu \, b({\mathbf{p_1}}) \nonumber \\
&\times&b^\dagger({\mathbf{p_2-k}})\, \gamma_\mu \, b({\mathbf{p_2}}) . 
\label{1-7}
\end{eqnarray}
Although this term contains only electron operators, its 
familiar interpretation is this: a gauge particle (the photon) 
with momentum ${\mathbf{k}}$ is emitted from particle 2 and absorbed 
by particle 1, causing transitions from ${\mathbf{p_2}}$ to 
${\mathbf{p_2-k}}$ and from ${\mathbf{p_1}}$ to ${\mathbf{p_1+k}}$. 
 
Mathematically, this term has a more prosaic interpretation: 
The $S$ matrix, when evaluated between incoming and outgoing 
states, describes a transition from an incoming 
two-particle product state to an entangled two-particle 
state and then back to an outgoing product state.
The entanglement is caused by the integration over ${\mathbf{k}}$, whereas the
integration over ${\mathbf{p_1}}$ and ${\mathbf{p_2}}$ means an integration
over a complete set of base states of the product state space.

\section{Two-particle state space and the fine-structure constant}

The functionality of the (first order) $S$ matrix, as just described, closely 
resembles the operation of a projection operator onto an intermediate 
two-particle subspace of the product state space.
In the following, this will be further substantiated.

Observe that the range of integration over ${\mathbf{p_1}}$ and 
${\mathbf{p_2}}$ is automatically restricted to the subspace of the parameter 
space with a total momentum ${\mathbf{P}}$, which equals the sum of the momenta 
of the incoming particles.
This means, the total momentum is conserved at each ``vertex''.
This property is preserved in higher orders of the perturbation series,
because these are obtained by iterating the first order $S$ matrix.
The entangled intermediate states, therefore, belong to a subspace of the 
product state space, characterized by a constant total momentum ${\mathbf{P}}$.
The fact that the states are entangled indicates a further restriction.
Since the perturbation algorithm is formulated in a covariant way, we can 
assume that this subspace is part of a relativistically invariant subspace,
characterized by $P^2 =$ some constant.
Let $\Omega$ be a manifold that parametrizes this subspace and let $V(\Omega)$ 
denote the volume of $\Omega$.

The states of this invariant subspace can be represented by linear combinations 
of base states 
$\left|{\mathbf{p_1}},{\mathbf{p_2}}\right>$, generated from the vacuum by two 
creation operators
\begin{equation}
\left|{\mathbf{p_1}},{\mathbf{p_2}}\right> =  \,b^\dagger({\mathbf{p_1}})\, 
b^\dagger({\mathbf{p_2}})\,\left| 0 \right> ,  \label{1-8}  
\end{equation}
with $(p_1 + p_2)^2 = P^2 = \mbox{constant}$.
The corresponding ``bra" states are
\begin{equation}
\left<{\mathbf{p_1}},{\mathbf{p_2}}\right| = \left< 0 \right| b({\mathbf{p_1}})\, 
b({\mathbf{p_2}}) . \label{1-8a}
\end{equation}
Observe, however, that by the anticommutation rule (\ref{1-4a}), these 
states are still normalized to the volume of the full product state space. 
Since a correct normalization is a precondition for the calculation of 
transition probabilities, the normalization has to be adjusted to the volume 
of this subspace.

Let us, for a while, forget that the volumes of the parameter spaces considered 
so far are infinite.
Then the correct normalization factor of a base state should 
be determined by the volume $V(\Omega)$, calculated from an embedding of $\Omega$
into the parameter space $\mathbb{R}^3 \times \mathbb{R}^3$ of the product state 
space, resulting in a factor $\sqrt{1/V(\Omega)}$.   
When these states or their creation/annihilation operators, respectively, 
are used to construct a projection operator such as integral (\ref{1-7}), then the 
normalization factor enters as $1/V(\Omega)$.

Since the goal of reverse engineering is a consistent mathematical 
description, we have to prove that this projection operator is in fact used 
in a way that is mathematically consistent with the requirements of a correct 
normalization.
Therefore our next step is, in general terms, to calculate $1/V(\Omega)$ and
then compare this value with a corresponding normalization factor that is
extracted from the perturbation algorithm.

$V(\Omega)$ can be determined independently from the evaluation of integral 
(\ref{1-7}) by calculating the integral over the manifold $\Omega$.
With the metric induced on $\Omega$ by its embedding into 
$\mathbb{R}^3 \times \mathbb{R}^3$, the integral can formally be 
written as
\begin{equation}
V(\Omega) = \int_\Omega \! d\lambda_{\mathbf{p_1},\mathbf{p_2}} , 
\label{1-8x}
\end{equation}
where $d\lambda_{\mathbf{p_1},\mathbf{p_2}}$ is the infinitesimal volume element on 
$\Omega$.

Let us, after the calculation of $V(\Omega)$, replace the volume element by
\begin{equation} 
d\omega_{\mathbf{p_1},\mathbf{p_2}} =
\frac{1}{V(\Omega)}\, d\lambda_{\mathbf{p_1},\mathbf{p_2}} ,
\label{1-8cc}
\end{equation}  
and then convert $d\omega_{\mathbf{p_1},\mathbf{p_2}}$ into a 
normalized Cartesian volume element 
\begin{equation} 
d\omega_{\mathbf{p_1},\mathbf{p_2}} =
\omega^2 \, d^5{\mathbf{p}},
\label{1-8c}
\end{equation}  
where $d^5{\mathbf{p}}$ is the Cartesian infinitesimal volume element, induced 
on $\Omega$ by its embedding into $\mathbb{R}^3 \times \mathbb{R}^3$, and
\begin{equation}
\omega^2 = \frac{d\omega_{\mathbf{p_1},\mathbf{p_2}}} 
{d^5{\mathbf{p}}}.
\label{1-8d}
\end{equation}
Then integral (\ref{1-8x}) takes on the form  
\begin{equation}
\int_\Omega \! d\omega_{\mathbf{p_1},\mathbf{p_2}} = 
\int_\Omega \! \omega^2 \, d^5{\mathbf{p}} = 1 . 
\label{1-8bb}
\end{equation}

The way in which $\omega^2$ is presented in equation (\ref{1-8d}) indicates
that only the ratio of the infinitesimal volume element
$d\omega_{\mathbf{p_1},\mathbf{p_2}}$ to the infinitesimal volume 
element $d^5{\mathbf{p}}$ needs to be determined.
Therefore, we are free to map both parameter spaces onto, for example, 
a finite (bounded) parameter space, before we perform the 
calculation of $\omega^2$, provided that this mapping does not change the 
ratio of the infinitesimal volume elements.

Based on equation (\ref{1-8d}), $\omega^2$ can be understood as a weight factor
that weights the contribution of a single product state to an irreducible 
two-particle state.
In the following, we will therefore refer to $\omega^2$ as ``weight factor''. 
Because of the relativistic covariance of the $S$ matrix, $\omega^2$ does not 
depend on the frame of reference.

After having calculated $\omega^2$, we will try to insert $\omega^2$ into 
integral (\ref{1-7}), to give this expression the consistent structure of a 
projection operator.
However, when inserting $\omega^2$, we notice that in the same position, the 
square of the empirical electromagnetic coupling constant $e$, i.e., the 
fine-structure constant $\alpha$, is also inserted ``by hand'' to reproduce the 
experimental data.
Hence, after having inserted the empirical value of $\alpha$, we cannot, in 
addition, insert the calculated weight factor without affecting the calculated 
transition amplitudes.
This conflict is resolved if $\alpha$ and the weight factor $\omega^2$
associated with the two-electron state space are one and the same. 

Under this premise, the calculation of $\omega^2$ takes on an entirely new
significance:
We should be able to identify the correct two-particle state space of 
$e$--$e$ scattering by selecting a promising state space, calculating 
the numerical value of $\omega^2$, and comparing it to the experimentally 
determined value of $\alpha$.
If we find that the two coincide, i.e.,
\begin{equation}
\omega^2 = \alpha , \label{1-9}
\end{equation}
we can consider this as experimental evidence that we have found the 
correct two-particle state space.

Now let us see how this idea can be put into practice.

\section{Irreducible two-particle\\representation}

The smallest relativistic invariant subspace of the product state space is the
space of an irreducible two-particle representation of the Poincar\'e group.
It represents the quantum mechanically correct description of an isolated 
two-particle system.

Let $p_1 = (p_1^0,{\mathbf{p_1}})$ and $p_2 = (p_2^0,{\mathbf{p_2}})$ be the 
4-momenta of two electrons.
They satisfy the mass shell relations
\begin{equation}
{p_1}^2 = {p_2}^2 = m^2 , \label{1-20}
\end{equation}               
where $m$ is the mass of the electron.
We also introduce the total and relative momentum by
\begin{equation}
P = p_1 + p_2 \; \mbox{ and } \; q = p_1 - p_2 . \label{1-21}
\end{equation}
By this definition, $P$ and $q$ satisfy 
\begin{equation}
P\,q = 0 . \label{1-22}
\end{equation}
Based on relation (\ref{1-22}), any two-particle state (reducible or irreducible) 
can be described by a total momentum $P$ and a spacelike momentum $q$, 
perpendicular to the timelike vector $P$. 
Perpendicular to a timelike vector means that $q$ is allowed to rotate by
the action of a SO(3) subgroup of the Lorentz group.

For an irreducible two-particle representation, the relation
\begin{equation}
P^2 = M^2 \label{1-23}   
\end{equation}
(mass hyperboloid) holds.
The ``mass'' $M$ corresponds to the value of one of two Casimir operators
(see below) that characterize an irreducible two-particle representation of 
the Poincar\'e group.
From equation (\ref{1-23}) we obtain
\begin{equation}
q^2 = 4m^2 - M^2 \le 0 . \label{1-24}
\end{equation} 
Equations (\ref{1-23}) and (\ref{1-24}) can be combined to
\begin{equation}
P^2 + q^2 = 4 m^2 . \label{1-27}
\end{equation}
Equation (\ref{1-24}) can be rewritten as
\begin{equation}
2 p_1 p_2 = M^2 - 2m^2     \label{1-25}
\end{equation}
or
\begin{equation}
2 p_1 P = 2 p_2 P = M^2  .  \label{1-26}
\end{equation}
Equations (\ref{1-25}) and (\ref{1-26}) correlate the particle momenta by fixing the 
angle between them and with respect to $P$.
Provided that $P$ is not in its rest frame, rotations with rotational 
axis $P$ preserve these angles. 
Since these rotations leave $P$ invariant, they can be related to a
rotational degree of freedom that is independent of the kinematics of $P$.
These rotations are described by an action of SO(2), acting synchronously on 
$p_1$ and $p_2$ and therefore also on the relative momentum $q$.
For $P$ in its rest frame, $P=(M,\mathbf{0})$, the orientation of the 
axis of the SO(2) rotations is undetermined, which allows for any axis 
perpendicular to $\mathbf{p_1} = -\mathbf{p_2}$.

Within an irreducible representation, the relative momentum $q$ can 
therefore be understood as a (2+1)-dimensional vector embedded in 
$\mathbb{R}^{3+1}$.

The action of SO(3,1) on $P$, together with the action of SO(2) on $q$,
generates the manifold $\Omega$, which parametrizes the state space of an 
irreducible two-particle representation labeled by $M$. 
The SO(2) moves within SO(3,1) as $P$ moves through the hyperboloid 
(\ref{1-23}).
The manifold $\Omega$ can therefore be described as a circle bundle 
over a hyperboloid.

\section{Calculation of the weight\\factor}

To determine the numerical value of $\omega^2$ for an irreducible two-particle
representation, we will evaluate integral (\ref{1-8bb}) 
``from scratch,'' using a bounded parametrization of $\Omega$, to take advantage
of the finite environment.

Due to the hyperbolic/circular structure of $\Omega$, we can expect 
that $\omega^2$ will contain contributions of volumes of circular or 
hyperbolic shapes.
This should remind us of a finding of the Swiss mathematician Armand Wyler,
who in 1971 published a formula that approxi\-mates the electromagnetic 
fine-structure constant $\alpha$ to a high degree of precision \cite{wyl71}. 
When Wyler found his formula, his favorite subject was:
``the various components of the boundaries of complex domains associated 
with Lie groups'' \cite{gil71}.
He observed that an expression, derived from the volumes of some homogeneous 
domains, related to Poisson's equation, delivered the numerical value of the 
fine-structure constant.
He published his finding in the hope that ``if he piqued the interest of 
the physics community, there might be more study of his favorite 
subject'' \cite{gil71}.
Unfortunately, the physics community neither understood his intention nor
his mathe\-matics.
Since Wyler was not able to put his observation into a convincing 
physical context, his paper was criticized \cite{rob71} and, in the 
following decades, it was considered as fruitless numerology \cite{gro89}.

Our calculation of $\omega^2$ will show that Wyler was perfectly right when he 
proposed his formula.
Just like Wyler, we will make use of some elements of the mathematical theory 
of symmetric homogeneous (bounded) domains (cf.\,\,e.g.\,\cite{vin12}). 

We can understand a symmetric homogeneous domain as an abstract parameter 
space on which a Lie group acts transitively as a symmetry group. 
``Transitively'' means that all points of the homogeneous domain can be obtained 
from any given point by an action of the symmetry group.
Accordingly, a quantum mechanical state space that has been parametrized 
by a symmetric homogeneous domain can be generated from a given point of
the domain by the simultaneous application of the full symmetry group to both 
the parameter space and the state space.
Thereby a one-to-one relation between the parameter space and the state space is
established.
This makes homogeneous domains an easy to handle tool for dealing with the 
corresponding state spaces.

The form of equation (\ref{1-27}), together with relation (\ref{1-22}), suggests 
a combination of $P$ with the (2+1)-dimensional $q$ to a (5+2)-dimensional 
vector $u$, by identifying $u_0=P_0,$ $u'_0=q_0$, 
$(u_1,u_2,u_3)=(P_1,P_2,P_3)$, $(u_4,u_5)=(q_1,q_2)$. 
Equation (\ref{1-27}) then becomes
\begin{equation}
u_0^2 + {u'}_0^2 - u_1^2 - u_2^2 - u_3^2 - u_4^2 - u_5^2 = 4 m^2 .
\label{2-1}
\end{equation}
This expression has the form of a ``mass hyperboloid'' with an SO(5,2) 
symmetry.
However, we have to keep in mind that on the hyperboloid (\ref{2-1}) there are no symmetry 
operations that ``rotate'' a spatial component of $P$ into a spatial component  of $q$.
So the values of $P^2$ and $q^2$ are separately kept constant under all
(permitted) symmetry operations.

Nevertheless, we can obtain rotations of spatial components of $P$ into 
such of $q$, provided that the timelike components $P_0$ and $q_0$ are  
automatically adjusted.
Then the values of $P^2$ and $q^2$ are again separately kept constant.
We will take advantage of this possibility below.

Considered as a hyperboloid with full SO(5,2) symmetry, the domain 
(\ref{2-1}) is isomorphic to the quotient group 
$\widehat{\Omega}$ = SO(5,2)/(SO(5)$\times$SO(2)), which
is a homogeneous domain with a transitive action of SO(5,2). 
With the group actions of the full SO(5,2), (\ref{2-1}) is an unbounded 
realization of the abstract manifold $\widehat{\Omega}$.
If we restrict the group action to SO(3,1) and SO(2), then (\ref{2-1}) is an 
unbounded realization of our parameter manifold $\Omega$. 

A well-known bounded realization of the homogeneous domain 
$\widehat{\Omega}$ is the complex {\it Lie ball} \cite{wyl71,hua63}
\begin{equation}
D^5 = \{z \in \mathbb{C}^5; 1 + |zz'|^2-2\bar{z}z' > 0, |zz'| < 1 \}. 
\label{2-4}
\end{equation}
The boundary of $D^5$ is given by
\begin{equation}
Q^5 = \{\xi = x\,e^{i\theta}; x \in \mathbb{R}^5, xx'=1 \},\;
0<\theta<\pi.                                   \label{2-6}
\end{equation}
(The vector $z'$ is the transpose of $z$, $\bar{z}$ is the complex conjugate 
of $z$.)
The Lie ball is included in the complex unit ball
\begin{equation}
C^5 = \{z \in \mathbb{C}^5; |zz'| < 1 \} \,\label{2-6a}
\end{equation}
and contains the real unit ball
\begin{equation}
B^5 = \{x \in \mathbb{R}^5; xx' < 1 \} . \label{2-6b}
\end{equation}

The complex unit ball is isomorphic to the upper half-space of $\mathbb{C}^5$,
whereas the Lie ball is isomorphic to the forward light cone in $5 + 2$
dimensions. 

There is some similarity to the mapping of the (unbounded) complex plane into 
the (bounded) Riemann sphere by a M\"obius transformation.
M\"obius transformations are conformal transformations. 
They leave invariant the form of volume elements but they change their sizes.
Whereas a subdomain of the complex plane may have an infinite volume, the 
volume of its image in the Riemann sphere is finite.
The Riemann sphere without the image of ``infinity'' has the same 
non-compact topology as the complex plane, but is bounded. 
By adding the image of infinity, the Riemann sphere becomes compact 
(this is the compactification of the complex plane).
On the internet, a very instructive animation of the M\"obius transformation
can be found \cite{moebius}. 
Readers not familiar with M\"obius transformations or the Riemann sphere may 
want to load the video of this animation before continuing.

Since the unbounded as well as the bounded realizations are true
realizations of $\widehat{\Omega}$, they are isomorphic.
Both can be used to parametrize a (fictive) $SO(5,2)$-invariant state space, 
but the bounded realization $D^5$ of $\widehat{\Omega}$ has the advantage 
that it provides a finite environment for calculating the integral 
$(\ref{1-8bb})$.
Therefore, the following evaluation of this integral will be based 
on the bounded realization of $\widehat{\Omega}$.

We can separate the integral into a spherical integral over the 4-dimensional 
surface $Q^5$, followed by a second integral over the radial direction of $D^5$. 
The spherical part of the integral is then given by 
\begin{equation}
\int_{Q^5} d^4x .      \label{2-9}
\end{equation}
The normalization of this integral requires the factor $1/V(Q^5)$, where
$V(Q^5)$ is the volume of $Q^5$.
This delivers a first contribution of $1/V(Q^5)$ to $\omega^2$.

Next we have to add the integration in the radial direction of $D^5$.
As indicated above, we want to obtain the infinitesimal volume element as
a Cartesian volume element.
Mapping a spherical volume to a rectangular one includes a step that is
known as the ``quadrature of the circle''. 
(As an example: the volume of the unit ball in three dimensions equals the 
volume of a cube with edge length $\sqrt[3]{4\pi/3}$.)

Consider the formula that relates the volume of a Lie ball $D^5_R$ with 
radius $R$ to the volume of the unit Lie ball $D^5$
\begin{equation}
V(D^5_R) = R^5 \, V(D^5) . \label{2-10a}
\end{equation}
When we project the volume of $D^5$ onto the real ball $B^5$ with  
surface $S^4$, then $V(D^5_R)$ can be expressed by the integral  
\begin{eqnarray}
&V(D^5_R)& = 5 \int^R_0 dr \int_{S^4} r^4 \,v^4\, d\omega_x , \nonumber \\
&\mbox{with}&\!\!\!\!v = V(D^5)^{\frac{1}{4}} , 
\mbox{  if } \int_{S^4} \! d\omega_x = 1 . \label{2-10b}
\end{eqnarray}
A rectangular volume with the same numerical value is given by 
\begin{equation}
5 \int^R_0 dr 
\int^r_0\! v\, dx_1
\int^r_0\! v\, dx_2
\int^r_0\! v\, dx_3
\int^r_0\! v\, dx_4  . \label{2-10c}
\end{equation}
This integral is an analogue to the ``quadrature of the circle''.
Unfortunately, it maps the volume of the Lie ball not to a cube, 
but to the cuboid
\begin{equation}
R^5 \times (1 \times v \times v \times v \times v)  .
\label{2-10d}
\end{equation}
The infinitesimal volume element $dr d\omega_x$ of integral (\ref{2-10b}) 
(e.g.\,\,at $r,x=1,0$) is accordingly mapped to the infinitesimal
volume element of integral (\ref{2-10c})
\begin{equation} 
dr\, v\, dx_1\, v\, dx_2\, v\, dx_3\, v\, dx_4  . \label{2-10e}
\end{equation}
Consequently, to obtain an isotropic volume element, the coordinate 
in the radial direction must be replaced (rescaled) according to
\begin{equation}
dr \rightarrow v\, dx_5 = V(D^5)^\frac{1}{4}\, dx_5  .  \label{2-11}
\end{equation} 
Therefore, to extend the 4-dimensional volume element $d^4x$ to a 
five-dimensional Cartesian isotropic volume element $d^5x$, we have to 
multiply $d^4x$ by the right hand side of relation (\ref{2-11}).
This adds a factor of $V(D^5)^\frac{1}{4}$ to $\omega^2$.

The fifth dimension also adds a factor to the normalization of the projection
operator, but for the Lie ball of radius $1$ this factor is equal to $1$, as 
can be seen by inspection of integral (\ref{2-10c}).

The infinitesimal volume element now refers to the full SO(5,2)-symmetric 
manifold $\widehat{\Omega}$, but remember that the original manifold $\Omega$ 
is subspace of $\widehat{\Omega}$ that is generated by rotations around 
four rotational axes instead of five. 
Therefore, the volume of $\Omega$ is smaller by a factor equal to the volume
of the quotient group SO(5)/SO(4), which is isomorphic to the real unit sphere 
$S^4$ in five dimensions (cf.\,\,e.g.\,\,\cite{sha12}).
Hence,
\begin{equation}
V(\Omega) = V(\widehat{\Omega})/V(S^4) . \label{2-11x}
\end{equation}

However, there is no indication that the perturbation algorithm excludes the 
integration over the direction of $x_5$.
Therefore, we cannot do other than keep this integration, together with the
corresponding normalization volume $V(\widehat{\Omega})$.
Keeping the five dimensions of the volume element means that on $\Omega$ we 
integrate through the $P$--$q$ boundary (on an integration path that connects 
$P$ with $q$).
Thereby we add up more points of the parameter space than the one-to-one 
relation between the parameter space and the state space allows.
But, as indicated above, these additional point are valid parameter 
combinations.
If we perform the same five-dimensional integration in the $S$ matrix element
(\ref{1-7}), we add up multiple copies of states, with multiplicity 
given by the volume of SO(5)/SO(4). 
We can compensate for the extra copies by simply adjusting the normalization 
of each state by a common factor and include this factor beforehand in the 
infinitesimal volume element.
This adds a factor of $1/V(S^4)$ to $\omega^2$.

This is a trick that works well with a projection operator that 
integrates with equal weights over the full parameter space.
But it conceals the fact that we are evaluating a five-dimensional integral 
over a basically four-dimensional manifold.
This discloses an inherent weakness in the perturbation algorithm of QED, 
which becomes obvious when the first order term is iterated:
The evaluation of higher order terms involves contractions (permutations) of
creation and annihilation operators. 
Thereby the structure of the projection operator gets lost and may become 
replaced by one of the notorious divergent loop structures of QED.  
Then the extra integration through the $P$--$q$ boundary cannot be compensated 
for as easily as before.
It becomes visible as an extra degree of freedom, leading to ill-defined 
integrals, which call for another trick to ``regularize'' them.
(A regularization method, based on distribution theory, can be found in Scharf
\cite{sch89}).
The insight into the mechanism that may lead to these divergences points out
a way to solve the divergence problem right at its source -- but that means
re-engineering the perturbation algorithm, which is not the subject of this
paper.

When we replace the total and relative momentum by the individual 
particle momenta $p_1$ and $p_2$, the Jacobian 
\begin{equation}
\frac{\partial(P,q)}{\partial(p_1,p_2)}      \label{2-11a}
\end{equation}
contributes a factor of 2 to the infinitesimal volume element.

We add a factor of $4\pi$, which is the volume of the unit sphere $S^2$
in three dimensions. 
It stands for integration over the full spatial angle.
(It is the same factor that enters the Poisson kernel in electrostatics.)

Collecting all factors results in a total weight factor $\omega^2$ of
\begin{equation}
8\pi \,V(D^5)^{\frac{1}{4}} \, / \, (V(S^4) \, V(Q^5))  .  \label{2-12}
\end{equation}
Expression (\ref{2-12}) is identical to Wyler's semi-empirical formula, which
here has been derived by reverse engineering the perturbation algorithm of QED.

Finally, we map the normalized volume element, constructed on the bounded 
realization, into the unbounded realization by a stereographic projection 
$T: x \rightarrow p$,
\begin{equation}
p =  \left( \frac{x_1}{1-x_5},\,\frac{x_2}{1-x_5},\,
\frac{x_3}{1-x_5},\,\frac{x_4}{1-x_5},\,\frac{x_5}{1-x_5} \right).
\label{2-15}
\end{equation}
The transformation (\ref{2-15}) is a conformal mapping.
The proof is by writing down (\ref{2-15}) for an infinitesimal cube.
Therefore, the isotropic volume element $\omega^2d^5x$ is mapped onto the 
isotropic Cartesian volume element $\omega^2d^5p$ in $\mathbb{R}^5$.
The value of $\omega^2$ is not touched by this mapping.
(A more intuitive, though less elegant, way would be to replace the unit Lie 
ball by a Lie ball with radius $R$ and then let $R \rightarrow \infty$.)
 
The volumes $V(D^5)$ and $V(Q^5)$ in Wyler's formula (\ref{2-12}) have been 
calculated by Hua \cite{hua63}. 
With
\begin{eqnarray}
V(Q^5) &=& \frac{8 \pi^3}{3}          	\label{2-16} \\
V(D^5) &=& \frac{\pi^5}{2^4\, 5!}     	\label{2-17}  \\
V(S^4) &=& \frac{8 \pi^2}{3}         	  \label{2-18}
\end{eqnarray}
we obtain 
\begin{equation}
\frac{9}{8 \pi^4} \! \left(\frac{\pi^5}{2^4 \, 5!}\right)^{1/4}  \! \!	
= \frac{9}{16 \pi^3} \! \left(\frac{\pi}{120}\right)^{1/4}   \! \!
= 1/137.03608245.   			          	\label{2-19}
\end{equation}
This value agrees up to a factor of $0.9999995$ with the
experimental (low energy) value of $\alpha$, which is 
the reciprocal of 137.035999084(51) \cite{han08}.

Note that, while deriving Wyler's formula, we have not touched the integrand
of integral (\ref{1-7}), which contains the ``physics'' of the $S$ matrix.
The whole calculation was based on the geometrical properties of the parameter
space only, without any direct involvement of the state space.
The operations on the parameter space, especially the mapping onto a finite
domain and back onto the infinite momentum space, followed transparent 
mathematical rules.
Therefore, we can be sure that we did not inadvertently modify the physical 
contents of the $S$ matrix. 

The extremely close agreement of $\omega^2$ with the (low energy) empirical 
value of $\alpha$ is a strong experimental indication that 
the (low-energy) ``physical'' two-particle state space of elastic 
$e$--$e$ scattering in fact matches an irreducible two-particle representation 
(of identical, massive, spin-1/2 particles) of the Poincar\'e group.
Since Joos's paper \cite{jos62} on the representations of the Lorentz group, 
these representations have been generally known.

Moreover, the numerical value of $\alpha$ can be regarded as a kind of 
checksum that double-checks the decisive steps of the reverse engineering 
procedure presented above. 
In fact, the individual elements of Wyler's formula helped the author 
more than once to avoid dead ends.  

The volume element on $\Omega$ still has only five dimensions, compared to 
six for the volume element $d{\mathbf{p_1}}d{\mathbf{p_2}}$ in the expression 
(\ref{1-7}) of the $S$ matrix. 
This shortfall can easily be resolved, without affecting the $S$ matrix, by 
simply extending the volume element of $\Omega$ to a six-dimensional 
one. This is because, in a two-particle scattering process, we can always 
orient the reference frame in such a way that the sixth momentum component 
of the incoming state is identically zero.
So a six-dimensional volume element in (\ref{1-7}) has only the ``cosmetic'' 
advantage of making the $S$ matrix look explicitly covariant.

Wyler's formula defines a geometrical factor that relates an irre\-ducible 
two-particle representation of the Poincar\'e group to a two-particle product 
representation, just as $\pi$ relates the circumference of a circle to its 
diameter.
In relating this geometrical factor to the empirical fine-structure constant,
we have to keep in mind that the latter is determined experimentally.
Therefore, all orders of the perturbation series, including non-elastic 
processes, contribute to its value.
The accumulation of these contributions is described by the 
renormalization group. 
This leads to a weakly energy dependent ``effective'' coupling constant 
-- the ``running coupling constant''.
At low energies, and depending on the experimental setup, non-elastic 
contributions of ``infrared photons'' can be kept well under control.
Therefore, the fine-structure constant measured by low-energy 
$e$--$e$ scattering comes close to the calculated value of the coupling
constant for elastic scattering.
This explains the success of Wyler's formula in reproducing the empirical 
value of $\alpha$.

\section{Angular momentum and\\entanglement}

Although we have identified the two-particle state space as an irreducible
representation of the Poincar\'e group, it is not yet clear why the
intermediate states in the $S$ matrix are entangled.
What explains the obvious absence of simple (separable) product states in the 
intermediate states?

Remember that we have based the calculation of $\omega^2$ on the observation 
that there is an internal rotational degree of freedom.
This corresponds to the orbital angular momentum of a two-particle configuration.

Irreducible representations of the Poincar\'e group are characterized by 
eigenvalues of the invariant (Casimir) operators (see e.g.\,\,Schweber \cite{sss}) 
\begin{equation}
P = p^\mu p_\mu
\end{equation}
and
\begin{equation}
W = -w^\mu w_\mu\;,\;\; \mbox{with} \;\; 
w_\sigma = \frac{1}{2} \epsilon_{\sigma \mu \nu \lambda} 
M^{\mu \nu}p^\lambda .
\end{equation}
Here $p^\mu$ and $M^{\mu \nu}$ are the operators of four-momentum and 
four-dimensional angular momentum, respectively.

To define a basis of the state space of an irreducible representation, we 
have to select a complete commuting set of operators.
Such a set consists of the momentum operators $p_\mu$ and one of the components
of $w_\mu$, say $w_p$, the component of the angular momentum operator in the 
direction of $p_\mu$.
The states of this basis can then be labeled by the quantum numbers of $p_\mu$ 
and $w_p$. 
$w_p$ is the generator of rotations with $p_\mu$ as the rotational axis. 
To give a two-particle state the property of an eigenstate of $w_p$, it is 
required that this state be a linear combination of all (pure) product 
states that can be reached from a given product state by such a rotation.
This necessarily gives a two-particle base state an entangled structure.
(Therefore, the separable states (\ref{1-8}), although used to generate 
irreducible two-particle states, do not form a basis of an irreducible 
two-particle state space.)

Entanglement correlates the individual particle states within the 
two-particle state. 
Obviously, it is this correlation that is observed as electromagnetic 
interaction.

\section{Vector potential}

In Feynman's formulation of the perturbation algorithm, the  
electromagnetic field operators have a surprisingly marginal role. 
In fact, Feynman deliberately eliminated these operators from the algorithm, 
to formulate it ``as a description of a direct interaction at a distance 
(albeit delayed in time) between charges'' \cite{rf2}.
This underlines the auxiliary role of the vector potential within QED.

In setting up the perturbation algorithm, the Dirac equation of the free 
electron is modified by adding a ``quantized vector potential'' to the 
momentum, in the sense of a ``minimal coupling to the electromagnetic field''.
Within the perturbation algorithm, the vector potential then obviously has 
the sole task of generating entangled states from incoming states. 
After having accomplished this, it is eliminated.

Based on this simple functionality, the reverse engineering approach must
understand the quantized vector potential as a sophisticated mathematical tool 
with the following properties:\\
a) It modifies the Dirac equation by a ``bookkeeping'' operator that stands for 
the ``potential'' that the state $\psi(p)$ may be changed, to become again a 
solution of the Dirac equation, but with the momentum $p + k$.\\
b) This change becomes active when and only when the operator $a(k)$ 
encounters its counterpart $a^\dagger(k)$.\\
The intended(!)$\,$result is that within the perturbation algorithm, two 
(incoming) single-particle states are mapped onto an entangled two-particle 
state with the same total momentum as the incoming states.
In this way a quantum mechanical transition from an incoming separable 
product state to a state of the corresponding irreducible two-particle 
representation is described.

The fact that $a(k)$ enters as a ``perturbation'' to the momentum $p$ in the 
Dirac equation, rather than, e.g., to the $\gamma p$-term, explains
why the $S$ matrix contains $\gamma$-matrices, something which, in a projection 
operator, is somewhat unexpected.
The strict pursuit of this perturbation ansatz, necessarily places the 
$\gamma$-matrices in the $S$ matrix.
The details can be found in any good textbook on QED 
(see e.g.\,\,Schweber \cite{sss1}).

\section{Virtual particles, vacuum\\fluctuations, and all that}

Feynman coined the term ``virtual quantum'' in his 1949/50 papers. 
Later it was replaced by ``virtual particle''.
It corresponds to the c-number that is left when the creation and an 
annihilation operators of the same particle type are permuted.
In Feynman graphs, these c-numbers are represented by internal lines 
connecting two vertices.
In the momentum representation, these c-numbers are essentially 
$\delta$ functions that ensure momentum conservation between two vertices.

In evaluating $S$ matrix elements, Feynman used the commutation relations 
to shift the creation and annihilation operators through the expression of the
matrix element, until they hit the vacuum state and thereby annihilate 
themselves.
In higher orders of the perturbation series, this leads to more and more 
``virtual particles''. 

The notion of ``virtual particle'' has triggered speculations about the 
``physical'' nature of virtual particles. 
It has been tried to give virtual particles some reality by considering them 
as particles that have ``left their mass shell''.
It has even been argued that, because of Heisenberg's uncertainty principle, 
virtual particles may become ``real'' for short periods of time.
(Ignoring the fact that this principle refers to particles, not to $\delta$ 
functions.)  
Together with the conviction that QED is the prototype of a quantum 
field theory, such ideas, although unsubstantiated, have strongly influenced 
the way we still think about QED and particle physics in general. 
Thereby they have unfortunately clouded our view of the comparatively 
simple mathematics of the perturbation algorithm for more than six decades.

The foregoing analysis is fully in line with Feynman's original notion of a
virtual quantum, and it is evident that in a simple and transparent product 
state space there is no room for speculations about $\delta$ functions 
becoming particles, or ``physical particles'' being ``dressed'' by clouds of 
particle/antiparticle pairs ``created from the vacuum''. 

The ``vacuum state'' used in the Fock space formalism is a symbolic 
state that only in connection with creation operators acting on it has a
counterpart in physical reality.
By reverse engineering, we have found that the ``physical'' state space is 
nothing other than a two-particle subspace of the Fock space.
Therefore, in QED there is no ``physical'' vacuum other then the (symbolic) 
vacuum of the Fock space.

A last remark concerns ``vacuum fluctuations''. 
There are ``vacuum graphs,'' which have internal lines, but no external 
(incoming or outgoing) lines.
Attempts have been made to understand these graphs as manifestations of 
quantum mechanically caused ``vacuum fluctuations''. 
The mathematical contents of these graphs (in the momentum representation)
are essentially a product of $\delta$ functions, whose arguments are momenta.
Therefore, they provide us, if at all, with the insight that, even when no 
particles are present, the principle of momentum conservation is observed.

Regarding the wide-spread opinion that the Casimir effect ``proves'' the 
existence of vacuum fluctuations, the reader is referred to Jaffe's 
article \cite{rlj}.

\section{Higher orders}

Our analysis of QED has so far been based on the first order of the 
perturbation series. 
Higher orders are obtained by iterating the first order operator.
Therefore, they are mathematically completely determined by the properties 
of the lowest order.

The iteration process is inherent to every perturbation approach.
What is special about a system of fermions, is that the anticommutation
relations allow interchanging the creation and annihilation operators.
Feynman has taken advantage of this property to set up practicable rules
for evaluating $S$ matrix elements.
In higher orders, these rules lead to a large variety of topologically 
different Feynman graphs.
Some of them have been interpreted as ``virtual pair creation'' 
or ``vacuum polarization''.
It is evident from our analysis of the two-particle $S$ matrix
that intermediate states are nothing other than two-particle
states, which do not give space for any additional pairs of particles 
``created from the vacuum''. 
So these interpretations merely give certain topological properties of 
Feynman graphs catchy names.

\section{Discussion}

The reverse engineering approach has led us to more than just a 
description of the perturbation algorithm.
The new insights into its mathematics, gained in this way, allows 
calculating the electromagnetic coupling constant $\alpha$.
The close agreement of the calculated with the empirical value 
provides evidence that the disclosed mathematical structure indeed reflects 
physical reality -- more than current concepts of interacting fields do, 
which leave the values of coupling constants undetermined.  
It reveals that in the perturbation algorithm of quantum electrodynamics, 
the $S$ matrix has the function of a projection operator onto intermediate 
irreducible two-particle states, with $\alpha$ acting as a normalization 
factor for these states.

With this understanding of the mathematical structure of the $S$ matrix, 
we can say:
The $S$ matrix describes a transition from a separable product state of two 
incoming electrons (preparation) to an intermediate irreducible 
two-particle state (propagation) and then back to a separable product state 
of two outgoing electrons (analysis).

The formation of irreducible intermediate states can be understood as the 
mani\-festation of a general rule of relativistic quantum mechanics: 
{\it An isolated quantum mechanical system is described by an 
irreducible representation of the Poincar\'e group}.
Therefore, the physical effects described by the $S$ matrix can be fully 
explained by elementary principles of relativistic quantum mechanics.

Whereas in the traditional interpretation of QED, the entanglement of  
two-particle states is caused by an exchange of ``virtual gauge 
particles,'' it has been shown that entanglement is a natural property of 
the state space of an irreducible two-particle representation of the 
Poincar\'e group.
Since we have not touched the mathematical structure of QED, we have thereby 
traced back the gauge invariance structure of QED to basic rules of quantum 
mechanics and Poincar\'e invariance.
However, now gauge invariance goes together with a certain value of 
the coupling constant, and we are lucky enough that this value matches
the (low energy) value of the empirical fine-structure constant.

Wyler's work has been of crucial importance for the foregoing analysis, 
because it has guided the author to valuable mathematical tools that used 
to be outside the horizon of a theoretical physicist.
Therefore, some of the objections that in the past were raised against 
Wyler's mathematics should be commented on.
A major objection was that Wyler used certain bounded spaces with a radius 
equal to 1.
It was argued (Robertson \cite{rob71}), that ``there is no known reason for 
setting r = 1,'' and it was suspected that a different radius would yield 
a different value for $\alpha$.
Another point of criticism was that Wyler could not clearly specify how the 
fourth-root factor entered his calculation.

From the derivation of Wyler's formula presented here, it should be clear 
that it does not depend on the radius of the Lie sphere.
The reason is that by equation (\ref{1-8d}) the weight factor $\omega^2$ is 
defined as the quotient of two infinitesimal volume elements on the surface of the 
two-particle mass hyperboloid.
Whether we map these volume elements to a Lie sphere with radius 1 or any other
radius or do not map it at all, does not have any influence on this quotient.
Speaking generally, the volumes in Wyler's formula are not the outcome of  
the mapping onto the Lie sphere, but rather reflect the internal geometrical 
structure of the homogeneous domain SO(5,2)/(SO(5)$\times$SO(2)), which is 
independent of any mapping.
The fourth-root factor has been identified as a trivial conversion factor 
relating a spherical to a Cartesian volume element.

In an answer to Robertson's objections, Gilmore wrote \cite{gil72}:\\ 
``Wyler's work has pointed out that it is possible to map an unbounded 
physical domain -- the interior of the forward light cone -- onto the interior 
of a bounded domain on which there also exists a complex structure. 
This mapping should prove of immense calculational value in the future.''

\section{Conclusion}

The empirical value of $\alpha$ provides experimental evidence that 
the state space of two interacting electrons belongs to an irreducible 
two-particle representation of the Poincar\'e group.

The electromagnetic interaction can, therefore, be fully understood
within the framework of a ``free'' relativistic multi-particle quantum 
theory, without the need to postulate an interaction with a ``gauge 
field'' -- provided that a general rule of relativistic quantum mechanics
is observed: 
Isolated systems are described by irreducible representations of the 
Poincar\'e group.

\section*{Acknowledgements}

I would like to thank several unknown referees for their critics, which 
helped me to improve my presentation. 
Special thanks go to Freeman Dyson for having read a previous version 
of the manuscript, and to Armand Wyler for his encouraging comments.
Last not least, I am grateful to Werner Heisenberg, who more than forty
years ago granted me a postgraduate studentship of the Max Planck Society. 
During my stay at the Max Planck Institute for Physics and Astrophysics, 
the first ideas of this work came about \cite{wst}.

\end{document}